\pdfoutput=1
\documentclass[11pt,a4paper,twoside,openright,closeany,textany]{book}
\usepackage{makeidx}
\usepackage{graphicx}
\usepackage{paralist}
\usepackage{caption} 
\usepackage{amsmath}
\usepackage{amssymb}
\usepackage{mathtools}
\usepackage{fancyhdr}
\usepackage{multicol}
\usepackage{multirow}

\usepackage{mdframed}
\usepackage[usenames,dvipsnames,svgnames,table]{xcolor}
\usepackage{array}
\usepackage[%
  colorlinks=true,%
  linkcolor=black,%
  urlcolor=black,%
  citecolor=black%
]{hyperref}
\usepackage{nameref} 
\usepackage{ifthen} 
\usepackage[font={small,sl}]{caption}
\usepackage[authoryear]{natbib}
\renewcommand\bibname{References}
\newcommand{\mychapbib}{
  \addcontentsline{toc}{section}{\bibname}
  \bibliographystyle{natbib}
  \bibliography{strucbioinf}
}
\usepackage[colorinlistoftodos]{todonotes}
\usepackage{letltxmacro}

\usepackage{tocstyle} 
\usetocstyle{standard}
\settocfeature{raggedhook}{\raggedright}
\newcommand{\bibfont}{\small\raggedright}
\def\cite{\citep}

\LetLtxMacro{\oldTodo}{\todo}
\renewcommand{\todo}[2][]{\oldTodo[#1]{TODO: #2}}

\usepackage[normalem]{ulem}

\newcommand\inwish[1]{\oldTodo[inline,color=SkyBlue]{WISH: #1}}

\newboolean{onechapter}

\newcommand{\AF}[1][~]{K.\@#1Anton#1Feenstra}
\newcommand{\SA}[1][~]{Sanne#1Abeln}
\newcommand{\JH}[1][~]{Jaap#1Heringa}
\newcommand{\AJ}[1][~]{Annika#1Jacobsen}
\newcommand{\HM}[1][~]{Halima#1Mouhib}
\newcommand{\PB}[1][~]{Punto#1Bawono}
\newcommand{\BS}[1][~]{Bas#1Stringer}
\newcommand{\QH}[1][~]{Qingzhen#1Hou}
\newcommand{\JvG}[1][~]{Juami#1H.\@#1M.\@#1van#1Gils}

\newcommand{\MD}[1][~]{Maurits#1Dijkstra}
\newcommand{\AM}[1][~]{Ali#1May}

\newcommand{\EvD}[1][~]{Erik#1van#1Dijk}
\newcommand{\NB}[1][~]{\mbox{Nicola}#1\mbox{Bonzanni}}
\newcommand{\OI}[1][~]{Olga#1Ivanova}
\newcommand{\TK}[1][~]{Tim#1Kwakman}
\newcommand{\JB}[1][~]{\mbox{Jochem}#1\mbox{Bijlard}}
\newcommand{\RB}[1][~]{\mbox{Robbin}#1\mbox{Bouwmeester}}
\newcommand{\RH}[1][~]{\mbox{Reza}#1\mbox{Haydarlou}}

\newcommand{\JG}[1][~]{\mbox{Jose}#1\mbox{Gavald\'a-Garc\'ia}}
\newcommand{\HdF}[1][~]{\mbox{Hans}#1\mbox{de}#1\mbox{Ferrante}}
\newcommand{\JV}[1][~]{\mbox{Jocelyne}#1\mbox{Vreede}}
\newcommand{\MO}[1][~]{\mbox{Mascha}#1\mbox{Okounev}}
\newcommand{\DG}[1][~]{\mbox{Dea}#1\mbox{Gogishvili}}

\newcommand{\LH}[1][~]{\mbox{Laura}#1\mbox{Hoekstra}}

\newcommand{\AR}[1][~]{\mbox{Arri\"en}#1\mbox{Symon}#1\mbox{Rauh}}

\newcommand{\KW}[1][~]{\mbox{Katharina}#1\mbox{Waury}}
\newcommand{\HI}[1][~]{\mbox{Hugo}#1\mbox{van}#1\mbox{Ingen}}
\newcommand{\TL}[1][~]{\mbox{Ting}#1\mbox{Liu}}

\newcommand{\orcid}[1]{\href{https://orcid.org/#1}{\raisebox{-0.7ex}{\protect\includegraphics[height=3ex]{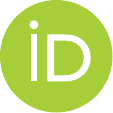}}}}
\definecolor{idgreen}{RGB}{166 206 57}
\newcommand{\mailid}[1]{\href{mailto:#1}{\raisebox{-0.3ex}{\color{idgreen}\textsf{\textbf{\Large \protect@}}}}}

\newcommand{\HMid}{\orcid{0000-0001-5031-3468}}
\newcommand{\JGid}{\orcid{0000-0001-6431-3442}}
\newcommand{\JvGid}{\orcid{0000-0003-3706-7818}}
\newcommand{\OIid}{\orcid{0000-0002-9111-4593}}

\newcommand{\DGid}{\orcid{0000-0001-8809-0861}}

\newcommand{\JVid}{\orcid{0000-0002-6977-6603}}
\newcommand{\RBid}{\orcid{0000-0001-6807-7029}}

\newcommand{\ARid}{\orcid{0000-0001-9707-3836}}

\newcommand{\KWid}{\orcid{0000-0002-8570-7640}}
\newcommand{\JHid}{\orcid{0000-0001-8641-4930}}
\newcommand{\HdFid}{\orcid{0000-0002-1772-0931}}
\newcommand{\BSid}{\orcid{0000-0001-7792-9385}}
\newcommand{\AJid}{\orcid{0000-0003-4818-2360}}
\newcommand{\EvDid}{\orcid{0000-0002-6272-2039}}
\newcommand{\QHid}{\orcid{0000-0002-9832-4518}}
\newcommand{\MDid}{\orcid{0000-0002-7971-6209}}
\newcommand{\HIid}{\orcid{0000-0002-0808-3811}}
\newcommand{\AMid}{\orcid{0000-0002-0551-9966}}
\newcommand{\LHid}{\orcid{0009-0003-4339-2470}}

\newcommand{\JBid}{\mailid{j.bijlard@gmail.com}}

\newcommand{\MOid}{\mailid{mascha.okounev@gmail.com}}
\newcommand{\PBid}{\mailid{puntobawono@gmail.com}}

\newcommand{\Angs}[1][~]{\text{\normalfont\AA}}

\renewcommand{\and}{\quad}

\newcommand{\pdbref}[1]{\href{http://www.rcsb.org/pdb/explore.do?structureId=#1}{PDB:#1}}
\newcommand{\arxiv}[2][UNDEFINED]{\href{https://arxiv.org/abs/#2}{\ifthenelse{\equal{#1}{UNDEFINED}}{arxiv.org/abs/#2}{#1}}}

\newcommand{\figref}[2][]{\hyperref[fig:#2]{Figure\@~\ref*{fig:#2}#1}}
\newcommand{\tabref}[1]{\hyperref[tab:#1]{Table \ref*{tab:#1}}}
\renewcommand{\eqref}[2][]{\hyperref[eq:#2]{Equation#1\@~\ref*{eq:#2}}}
\newcommand{\panelref}[2][]{%
    \ifthenelse{\boolean{onechapter}}{%
        \hyperref[panel:#2]{Panel\@~``\nameref{panel:#2}#1''}%
    }{%
        \hyperref[panel:#2]{Panel\@~\ref*{panel:#2}#1}%
    }%
}
\newcommand{\secref}[2][n]{%
    \hyperref[sec:#2]{%
        \ifthenelse{\equal{#1}{n} }{Section\@~\ref*{sec:#2}}{}% just number
        \ifthenelse{\equal{#1}{nn}}{Section\@~\ref*{sec:#2} ``\nameref{sec:#2}''}{}% nm & nr
        \ifthenelse{\equal{#1}{N} }{``\nameref{sec:#2}''}{}% just quoted name
        \ifthenelse{\equal{#1}{NN} }{\nameref{sec:#2}}{}% just name
    }%
}
\newcommand{\chref}[2][n]{%
    \ifthenelse{\boolean{onechapter}}{%
        \ifthenelse{\equal{#2}{ChPref}     }{\arxiv[Chapter \ref*{ch:#2} ``\nameref*{ch:#2}'']{1801.09442}}{}%
        \ifthenelse{\equal{#2}{ChIntroPS}  }{\arxiv[Chapter \ref*{ch:#2} ``\nameref*{ch:#2}'']{2307.02169}}{}%
        \ifthenelse{\equal{#2}{ChDetVal}   }{\arxiv[Chapter \ref*{ch:#2} ``\nameref*{ch:#2}'']{2108.02706}}{}%
        \ifthenelse{\equal{#2}{ChStrucAli} }{\arxiv[Chapter \ref*{ch:#2} ``\nameref*{ch:#2}'']{2307.02170}}{}%
        \ifthenelse{\equal{#2}{ChDBClass}  }{\arxiv[Chapter \ref*{ch:#2} ``\nameref*{ch:#2}'']{2307.02171}}{}%
        \ifthenelse{\equal{#2}{ChFunc}     }{\arxiv[Chapter \ref*{ch:#2} ``\nameref*{ch:#2}'']{1801.09442}}{}%
        \ifthenelse{\equal{#2}{ChIntroPred}}{\arxiv[Chapter \ref*{ch:#2} ``\nameref*{ch:#2}'']{1712.00407}}{}%
        \ifthenelse{\equal{#2}{ChHomMod}   }{\arxiv[Chapter \ref*{ch:#2} ``\nameref*{ch:#2}'']{1712.00425}}{}%
        \ifthenelse{\equal{#2}{ChSSPred}   }{\arxiv[Chapter \ref*{ch:#2} ``\nameref*{ch:#2}'']{1801.09442}}{}
        \ifthenelse{\equal{#2}{ChFuncPred} }{\arxiv[Chapter \ref*{ch:#2} ``\nameref*{ch:#2}'']{2307.02173}}{}%
        \ifthenelse{\equal{#2}{ChIntroDyn} }{\arxiv[Chapter \ref*{ch:#2} ``\nameref*{ch:#2}'']{2307.02174}}{}%
        \ifthenelse{\equal{#2}{ChThermo}   }{\arxiv[Chapter \ref*{ch:#2} ``\nameref*{ch:#2}'']{2307.02175}}{}%
        \ifthenelse{\equal{#2}{ChMD}       }{\arxiv[Chapter \ref*{ch:#2} ``\nameref*{ch:#2}'']{2307.02176}}{}%
        \ifthenelse{\equal{#2}{ChMC}       }{\arxiv[Chapter \ref*{ch:#2} ``\nameref*{ch:#2}'']{2307.02177}}{}%
    }{
    \hyperref[ch:#2]{%
        \ifthenelse{\equal{#1}{n} }{Chapter \ref*{ch:#2}}{}% just number
        \ifthenelse{\equal{#1}{nn}}{Chapter \ref*{ch:#2} ``\nameref{ch:#2}''}{}% name & number
        \ifthenelse{\equal{#1}{N} }{``\nameref{ch:#2}''}{}% just name
      }%
  }%
}
\newcommand{\chrefname}[1]{\hyperref[ch:#1]{Chapter \ref*{ch:#1} ``\nameref{ch:#1}''}}
\newcommand{\partref}[1]{\hyperref[#1]{Part \ref*{#1}}}
\newcommand{\appref}[1]{\hyperref[app:#1]{Appendix \ref*{app:#1}}}

\newcommand{\figsource}[1]{\protect\footnote{Figure source location: \url{#1}}}

\newcommand\blindfootnote[1]{%
  \begingroup
  \renewcommand\thefootnote{}\footnote{#1}%
  \addtocounter{footnote}{-1}%
  \endgroup
}

\makeatletter \@beginparpenalty=5000 \makeatother

\usepackage{listings}
\definecolor{backcolour}{rgb}{0.95,0.95,0.92}
\definecolor{codegreen}{rgb}{0,0.6,0}
\definecolor{codegray}{rgb}{0.5,0.5,0.5}
\definecolor{codered}{rgb}{0.8,0,0.0}
\definecolor{codeblue}{rgb}{0.0,0,0.8}
\lstdefinestyle{codeStyle}{
    backgroundcolor=\color{backcolour},   
    commentstyle=\color{codegreen},
    keywordstyle=\color{codeblue},
    numberstyle=\tiny\color{codegray},
    stringstyle=\color{codegray},
    numbers=left,                    
    tabsize=2
} 
\makeindex

\begin{document}

\setboolean{onechapter}{true}

\pagestyle{fancy}
\lhead[\small\thepage]{\small\sf\nouppercase\rightmark}
\rhead[\small\sf\nouppercase\leftmark]{\small\thepage}
\newcommand{\innerfoot}{\footnotesize{\sf{\copyright} Feenstra \& Abeln}, 2014-2023}
\newcommand{\outerfoot}{\footnotesize \sf Intro Prot Struc Bioinf}
\lfoot[\outerfoot]{\innerfoot}
\cfoot{}
\rfoot[\innerfoot]{\outerfoot}
\renewcommand{\footrulewidth}{\headrulewidth}

\mainmatter

\title{Introduction to Structural Bioinformatics}
\author{\AF
  \and \SA
  \and
  \\[10ex]
  \textrm{\footnotesize Centre for Integrative Bioinformatics (IBIVU)}\\
  \textrm{\footnotesize Department of Computer Science}\\
  \textrm{\footnotesize Vrije Universiteit, De Boelelaan 1081A, 1081 HV Amsterdam, Netherlands}
}

\maketitle

\section*{Abstract}
While many good textbooks are available on Protein Structure, Molecular Simulations, Thermodynamics and Bioinformatics methods in general, there is no good introductory level book for the field of Structural Bioinformatics. This book aims to give an introduction into Structural Bioinformatics, which is where the previous topics meet to explore three dimensional protein structures through computational analysis. We provide an overview of existing computational techniques, to validate, simulate, predict and analyse protein structures. More importantly, it will aim to provide practical knowledge about how and when to use such techniques. We will consider proteins from three major vantage points: Protein structure quantification, Protein structure prediction, and Protein simulation \& dynamics.
\newpage

\vfill

\mainmatter

\newpage
\setcounter{chapter}{-1}
\setcounter{chapter}{-1}
\ifthenelse{\boolean{onechapter}}{
\chapter*{Contents}
\newcommand{\nyp}{{\bf *} }
\begin{itemize}
    \item[\textsl{\textbf{Part I\quad Structure \& structure comparison}}] 
    \item[]
    \begin{itemize}
        \item[\bf\chref{ChIntroPS}]
        \item[\bf\chref{ChDetVal}] 
        \item[\bf\chref{ChStrucAli}] 
        \item[\bf\chref{ChDBClass}] 
    \end{itemize}\vspace*{-\baselineskip} \vspace*{\medskipamount}
    \item[\textsl{\textbf{Part II\quad Structure prediction}}] 
    \item[]
    \begin{itemize}
        \item[\bf\chref{ChIntroPred}]
        \item[\bf\chref{ChHomMod}]
        \item[\nyp \bf\chref{ChSSPred}] 
        \item[\bf\chref{ChFuncPred}] 
    \end{itemize}\vspace*{\medskipamount}
    \item[\textsl{\textbf{Part III\quad Simulation \& dynamics}}] 
    \item[]
    \begin{itemize}
        \item[\bf\chref{ChIntroDyn}] 
        \item[\bf\chref{ChThermo}] 
        \item[\bf\chref{ChMD}] 
        \item[\bf\chref{ChMC}] 
    \end{itemize}\vspace*{\medskipamount}
\end{itemize}
\noindent
{\footnotesize * Above links will bring you to the separately published chapters.
If you return back here, or if you arrived here following a link from one of the chapters, the chapter you are looking for has not yet been published.}
}{}

\chapter*{Preface}
\blindfootnote{\hspace{-1.5\parindent}\textsf{Available online at: \arxiv{1801.09442}}}
\label{ch:ChPref}

Why did we write this book? Firstly, because all the authors involved are very enthousiastic about protein structures and scientific computation, and secondly because we have been setting up a successful MSc level Bioinformatics in Structural Bioinformatics course that is in immediate need for a good text book. 

While many good textbooks are available on Protein Structure, Molecular Simulations, Thermodynamics and Bioinformatics methods in general, there is no good introductory level book for the field of Structural Bioinformatics. This book aims to give an introduction into Structural Bioinformatics, which is where the previous topics meet to explore three dimensional protein structures through computational analysis. 

\begin{figure}
  \centerline{
    \includegraphics[width=0.9\linewidth]{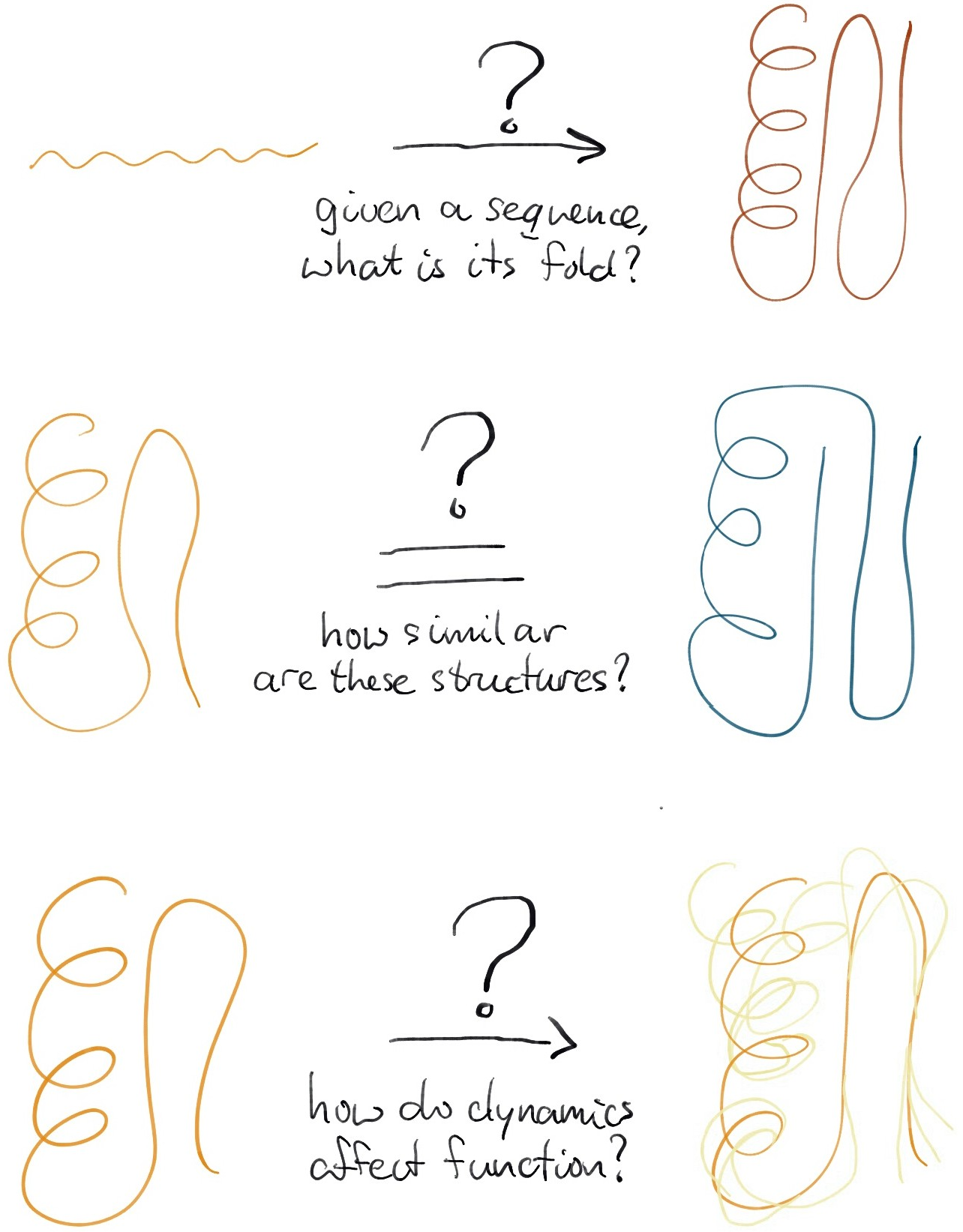}
  }
  \caption{Within the field of Structural Bioinformatics three dimensional protein structures are investigated through computational analysis. Important problems that may be addressed computationally are shown in the form of cartoons. Firstly, how does the genomic sequence of a gene translate into the folded, functional protein structure? Secondly, when considering two proteins, how similar are their structures? And, last but not least, since we know proteins are not static entities, how do flexibility and dynamics play a role in the function of the protein?}
  \label{fig:ChPref:Cartoons}
\end{figure}

This book will provide you an overview of existing computational techniques, to validate, simulate, predict and analyse protein structures. More importantly, it will aim to provide practical knowledge about how and when to use such techniques. We will consider proteins from three major vantage points, as illustrated in \figref{ChPref:Cartoons}: 
\begin{compactitem}
\item[\it I)] Protein structure \&  structure comparison; 
\item[\it II)] Protein structure prediction; and 
\item[\it III)] Protein simulation \& dynamics.
\end{compactitem}

\textsl{Part I Structure \& structure comparison} deals with comparing one protein structure to another. This can either be in a general sense: a protein structure is compared to a large set of reference structures to validate the experimental reliability of the structure. Or the comparison may be to one or more specific protein structures, with the question how similar the protein folds are.

\textsl{Part II Structure prediction} deals with the question, how to predict the structure given a protein sequence. We start with a graceful introduction to protein structure basics \citep{Abeln2017a}. We will see that there is a wide range of methods available, and that the reliability of such method varies strongly. It is important to understand how and when structure prediction will give you trustworthy outcomes \citep{Abeln2017b}. Then we reach perhaps the most salient question: from an available structure, what can we say about the possible function this protein could perform. 

\textsl{Part III Simulation \& dynamics} considers the dynamics or the ensemble of possible structures a protein may take up. Generally speaking however, it is not yet possibly to allow proteins to fold within a simulation, nevertheless simulations are vital to understand functional mechanisms of structural ensembles of proteins, specifically in context of small metabolites, the cell membrane or other proteins.

Herein, we introduce relevant concepts such as structure basics, structure prediction, homology modeling, statistical thermodynamics, simulation. From here, the reader may be able to acquire further information through self study and further references. In particular, we will treat the protein not as a rigid structural entity, but as dynamic ensembles of structures which make the protein able to perform its function or functions.

Hence this book aims to provide a framework for all important concepts within the field of Structural Bioinformatics. However, we do not aim to give extensive reviews of all the methods available in each of these subjects; for these we will refer, where appropriate, to other books.

The primary audience of this book are master students in the area of bioinformatics, or a related discipline like biotechnology. Basic background knowledge is assumed on protein structure, bioinformatics, sequence analysis, dynamic programming algorithms, calculus and basic chemistry. A few good books we would like to mention here, upfront, as they may be important reference material for readers of this book, depending on their background:
\begin{compactitem}
\item Protein Structure -- \citet{BrandenTooze}
\item Understanding Bioinformatics -- \citet{ZvelebilBaum}
\item Essential Bioinformatics -- \citet{Xiong} (very basic)
\item An introduction to Thermal Physics -- \citet{Schroeder}
\item Introduction to Python -- \citet{Lutz}
\end{compactitem}

And some more advanced books, that go further on particular topics than this book:
\begin{compactitem}
\item Structural Bioinformatics -- \citet{GuBourne}
\item Sequence Analysis -- \citet{Durbin}
\item Molecular Simulation -- \citet{FrenkelSmit}
\item Molecular Modelling -- \citet{Leach}
\end{compactitem}

We hope to provide the reader with knowledge about the type of problems that are scientifically feasible to solve. For example, predicting a protein structure through homology modeling will generally give high quality and reliable results. In contrast, we also would like to point out which problems are still unlikely to yield good results, such as predicting protein-protein interactions, and which are still out of scope altogether for existing methods, such as correct prediction of protein folding from sequence information alone. Most importantly, we will introduce the fundamental concepts on which the method are based, and what assumptions there are for using such methods. Therefore it should become apparent which inherent limitations the techniques have and what the techniques can successfully be used for.

\begin{figure}
  \centerline{
    \includegraphics[width=0.7\linewidth]{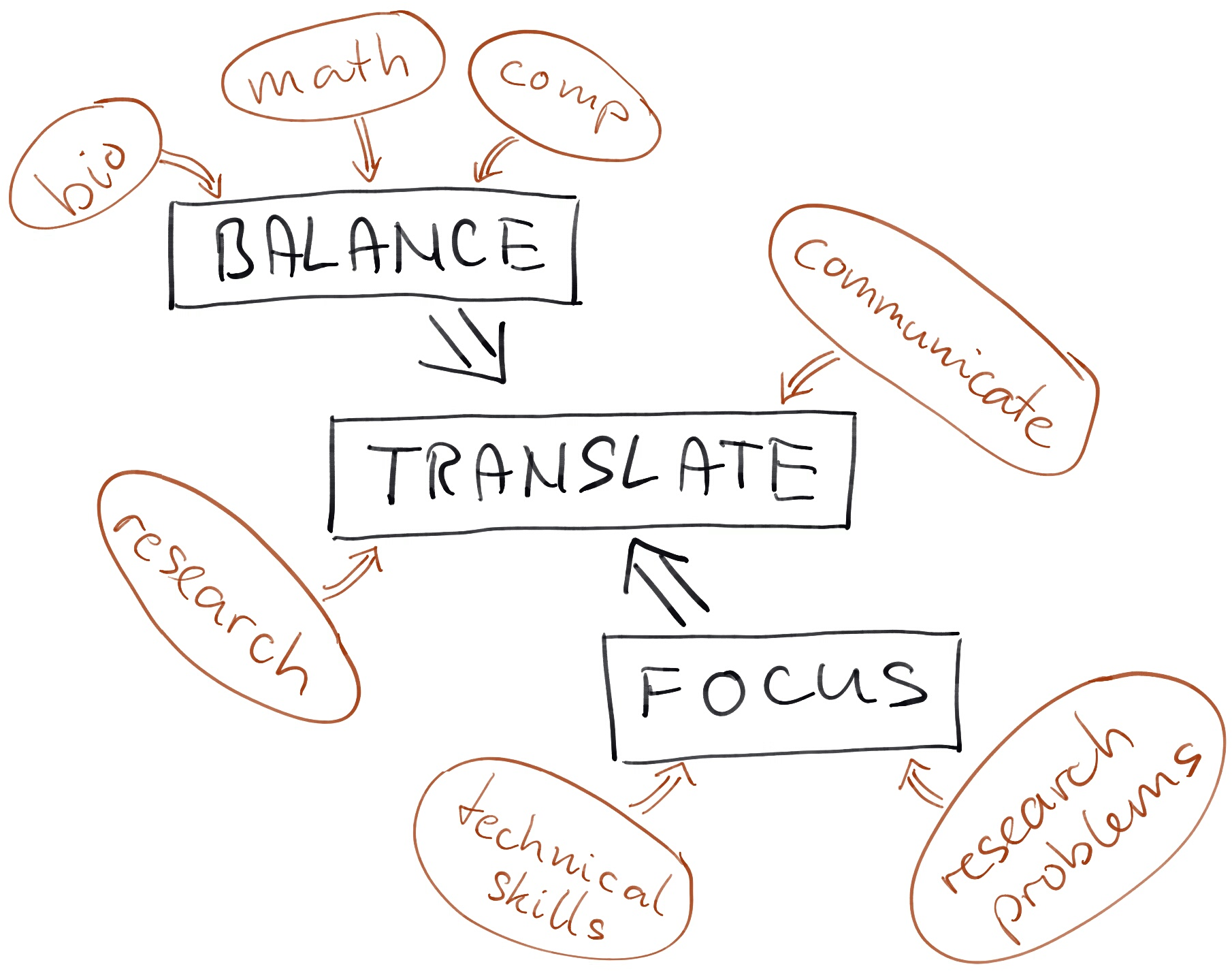}
  }
  \caption{Conceptual organisation of our bioinformatics education programme along three key elements: Translate, Balance, and Focus.}
  \label{fig:ChPref:concepts-btf}
\end{figure}

The setup of this books follows the same guiding principle that we apply throughout our masters' programme. We introduce a \textbf{focus} on open and challenging research problems to solve, and the technical skills to solve them; a \textbf{balance} between basic skills, such as biology, mathematics (modelling) and computational tools (programming); then, building on this focus and balance, \textbf{translate} between different adjoining disciplines, and between tools and methods on the one hand, and application and problems on the other \cite{Abeln2013,Feenstra2018}. This is summarized in \figref{ChPref:concepts-btf}.

Here, we would like tho thank all chapter authors: 
\AJ[ ]~\AJid, 
\AM[ ]~\AMid, 
\AR[ ]~\ARid, 
\BS[ ]~\BSid, 
\DG[ ]~\DGid, 
\EvD[ ]~\EvDid, 
\HdF[ ]~\HdFid, 
\HI[ ]~\HIid, 
\HM[ ]~\HMid, 
\JB[ ]~\JBid, 
\JG[ ]~\JGid, 
\JH[ ]~\JHid, 
\JvG[ ]~\JvGid, 
\JV[ ]~\JVid, 
\KW[ ]~\KWid, 
\LH[ ]~\LHid, 
\MD[ ]~\MDid, 
\MO[ ]~\MOid, 
\OI[ ]~\OIid, 
\PB[ ]~\PBid, 
\QH[ ]~\QHid, and 
\RB[ ]~\RBid.
Lastly, we would like to thank all our students who have followed the MSc course Structural Bioinformatics at the VU University in Amsterdam for their enthusiasm in pointing out mistakes in our lectures and asking important additional questions; without this vital input it would have been absolutely impossible to write this book. In particular, we thank %
\NB[ ], Ashley Gallagher, \TK[ ], \TL[ ], Arthur Goetzee, and \RH[ ]
for insightful discussions and critical proofreading of early versions. 

\ifthenelse{\boolean{onechapter}}{}{%
    \mychapbib
}

\mychapbib

\printindex


\begin{thebibliography}{}

\bibitem[Abeln {\em et~al.}(2013)Abeln, Molenaar, Feenstra, Hoefsloot, Teusink,
  and Heringa]{Abeln2013}
Abeln, S., Molenaar, D., Feenstra, K.~A., Hoefsloot, H. C.~J., Teusink, B., and
  Heringa, J. (2013).
\newblock {Bioinformatics and Systems Biology: bridging the gap between
  heterogeneous student backgrounds}.
\newblock {\em Briefings in Bioinformatics\/}, {\bf 14}(5), 589--598.

\bibitem[Abeln {\em et~al.}(2017a)Abeln, Heringa, and Feenstra]{Abeln2017a}
Abeln, S., Heringa, J., and Feenstra, K.~A. (2017a).
\newblock {Introduction to Protein Structure Prediction}.
\newblock {\em arXiv\/}, {\bf 1712.00407}.

\bibitem[Abeln {\em et~al.}(2017b)Abeln, Heringa, and Feenstra]{Abeln2017b}
Abeln, S., Heringa, J., and Feenstra, K.~A. (2017b).
\newblock {Strategies for protein structure model generation}.
\newblock {\em arXiv\/}, {\bf 1712.00425}.

\bibitem[Branden and Tooze(1998)Branden and Tooze]{BrandenTooze}
Branden, C. and Tooze, J. (1998).
\newblock {\em {Introduction to protein structure}\/}.
\newblock garland publishing, New York.

\bibitem[Durbin {\em et~al.}(1998)Durbin, Eddy, Krogh, and Mitchison]{Durbin}
Durbin, R., Eddy, S.~R., Krogh, A., and Mitchison, G. (1998).
\newblock {\em {Biological Sequence Analysis}\/}.
\newblock Cambridge University Press.

\bibitem[Feenstra {\em et~al.}(2018)Feenstra, Abeln, Westerhuis, {Brancos dos
  Santos}, Molenaar, Teusink, Hoefsloot, Heringa, Brancos, Molenaar, Teusink,
  Hoefsloot, and Heringa]{Feenstra2018}
Feenstra, K.~A., Abeln, S., Westerhuis, J.~A., {Brancos dos Santos}, F.,
  Molenaar, D., Teusink, B., Hoefsloot, H. C.~J., Heringa, J., Brancos, F.,
  Molenaar, D., Teusink, B., Hoefsloot, H. C.~J., and Heringa, J. (2018).
\newblock {Training for translation between disciplines : a philosophy for life
  and data sciences curricula}.
\newblock {\em Bioinformatics\/}, {\bf 34}(13), 1--9.

\bibitem[Frenkel and Smit(2002)Frenkel and Smit]{FrenkelSmit}
Frenkel, D. and Smit, B. (2002).
\newblock {\em {Understanding Molecular Simulation: From Algorithms to
  Applications}\/}, volume~1 of {\em Computational Science Series\/}.
\newblock Academic Pr, San Diego, second edition.

\bibitem[Gu and Bourne(2009)Gu and Bourne]{GuBourne}
Gu, J. and Bourne, P.~E. (2009).
\newblock {\em {Structural bioinformatics}\/}.
\newblock John Wiley {\&} Sons,, Hoboken, 2nd ed. nv edition.

\bibitem[Leach(2001)Leach]{Leach}
Leach, A. (2001).
\newblock {\em {Molecular Modelling: Principles and Applications}\/}.
\newblock Pearson.

\bibitem[Lutz(2013)Lutz]{Lutz}
Lutz, M. (2013).
\newblock {\em {Learning Python}\/}.
\newblock O'Reilly.

\bibitem[Schroeder(1999)Schroeder]{Schroeder}
Schroeder, D.~V. (1999).
\newblock {\em {An Introduction to Thermal Physics}\/}.
\newblock Addison-Wesley Publishing Company, San Francisco, CA.

\bibitem[Xiong(2006)Xiong]{Xiong}
Xiong, J. (2006).
\newblock {\em {Essential Bioinformatics}\/}.
\newblock Cambridge University Press.

\bibitem[Zvelebil and Baum(2008)Zvelebil and Baum]{ZvelebilBaum}
Zvelebil, M. and Baum, J. (2008).
\newblock {\em {Understanding Bioinformatics}\/}.
\newblock Garland Science, Taylor {\&} Francis Group, New York -- London.

\end{thebibliography}
\end{document}